\documentclass{eptcs}
\providecommand{\event}{CompMod 2009} 
\providecommand{\volume}{??}
\providecommand{\anno}{2009}
\providecommand{\firstpage}{1}
\providecommand{\eid}{??}
\usepackage{breakurl}        

\usepackage{graphicx}
\usepackage{color}
\usepackage{epsfig}
\usepackage{amssymb}
\usepackage{amsmath}
\usepackage{amsthm}
\usepackage[ansinew]{inputenc}
\usepackage{dsfont}
\usepackage{fancybox,tikz}

\newtheorem{proposition}{Proposition}
\newtheorem{definition}{Definition}

\newtheorem{example}{Example}
\newtheorem{remark}{Remark}

\title{Mutual Mobile Membranes with Timers\footnote{The research for this paper was partially supported by CNCSIS IDEI
402/2007 and CNCSIS TD 345/2008.}}
\author{Bogdan Aman
\institute{Romanian Academy, Institute of Computer Science}
\institute{A.I.Cuza University of Ia\c si, Romania}
\email{baman@iit.tuiasi.ro} \and Gabriel Ciobanu \institute{Romanian
Academy, Institute of Computer Science} \institute{A.I.Cuza
University of Ia\c si, Romania} \email{gabriel@info.uaic.ro} }
\def\titlerunning{Mutual Mobile Membranes with Timers}
\def\authorrunning{B. Aman and G. Ciobanu}
\begin{document}
\maketitle

\begin{abstract}
A feature of current membrane systems is the fact that objects and
membranes are persistent. However, this is not true in the real
world. In fact, cells and intracellular proteins have a well-defined
lifetime. Inspired from these biological facts, we define a model of
systems of mobile membranes in which each membrane and each object
has a timer representing their lifetime. We show that systems of
mutual mobile membranes with and without timers have the same
computational power. An encoding of timed safe mobile ambients into
systems of mutual mobile membranes with timers offers a relationship
between two formalisms used in describing biological systems.
\end{abstract}

\section{Introduction}
\label{section:introduction}

Membrane systems are essentially parallel and nondeterministic
computing models inspired by the compartments of eukaryotic cells
and their biochemical reactions. The structure of the cell is
represented by a set of hierarchically embedded regions, each one
delimited by a surrounding boundary (called membrane), and all of
them contained inside an external special membrane called {\it
skin}. The molecular species (ions, proteins, etc.) floating inside
cellular compartments are represented by multisets of objects
described by means of symbols or strings over a given alphabet. The
objects can be modified or communicated between adjacent
compartments. Chemical reactions are represented by evolution rules
which operate on the objects, as well as on the compartmentalized
structure (by dissolving, dividing, creating, or moving membranes).

A membrane system can perform computations in the following way:
starting from an initial configuration which is defined by the
multiset of objects initially placed inside the membranes, the
system evolves by applying the evolution rules of each membrane in a
nondeterministic and maximally parallel manner. A rule is applicable
when all the objects which appear in its left hand side are
available in the region where the rule is placed. The maximally
parallel way of using the rules means that in each step, in each
region of the system, we apply a maximal multiset of rules, namely a
multiset of rules such that no further rule can be added to the set.
A halting configuration is reached when no rule is applicable. The
result is represented by the number of objects from a specified
membrane.

Several variants of membrane systems are inspired by different
aspects of living cells (symport and antiport-based communication
through membranes, catalytic objects, membrane charge, etc.). Their
computing power and efficiency have been investigated using the
approaches of formal languages and grammars, register machines and
complexity theory. Membrane systems (also called P systems) are
presented together with many variants and examples in \cite{Paun02}.
Several applications of these systems are presented in \cite{vaps}.
An updated bibliography can be found at the P systems web page
\cite{ppage}.

A first attempt to define mobile P systems is presented in
\cite{Petre99-02} where the rules are similar to those of mobile
ambients \cite{Cardelli98}. Inspired by the operations of
endocytosis and exocytosis, namely moving a membrane inside a
neighbouring membrane (endocytosis) and moving a membrane outside
the membrane where it is placed (exocytosis), the P systems with
mobile membranes are introduced in \cite{KrishnaPaun05} as a variant
of P systems with active membranes \cite{Paun02}.

The systems of {\it mutual mobile membranes} represent a variant of
P systems with mobile membranes in which the endocytosis and
exocytosis work whenever the involved membranes ``agree'' on the
movement; this agreement is described by using dual objects $a$ and
$\overline{a}$ in the corresponding rules. The operations governing
the mobility of the systems of mutual mobile membranes are called
mutual endocytosis (mutual endo), and mutual exocytosis (mutual
exo).

The structure of the paper is as follows. In Section
\ref{section:mutual_mobile_membranes} we give a formal definition of
the new class of mutual mobile membranes together with their
biological motivation. Section
\ref{section:timed_mutual_mobile_membranes} contains the formal
definition of systems of mutual mobile membranes with timers, a
variant of systems of mutual mobile membranes in which timers are
attached to each object and each membrane. Section
\ref{section:tMM_MM} contains some results which show that we do not
obtain more computational power by adding timers to objects and
membranes into a system of mutual mobile membranes. Section
\ref{section:tMM_tMA} presents a translation of timed safe mobile
ambients into systems of mutual mobile membranes with timers.
Related work, conclusion and references finalize the~paper.

\section{Systems of Mutual Mobile Membranes}
\label{section:mutual_mobile_membranes}

Endocytosis and exocytosis are general terms which refer to the
process by which anything is taken into or expelled from the cell
through the action of vacuoles. Exocytosis involves the movement of
materials out of the cytoplasm of the cell using ATP energy. In
exocytosis, a vesicle (vacuole) migrates to the membrane inner
surface and fuses with the cell membrane. This process of exocytosis
is how the cells of glands producing proteins (enzyme and steroids)
export molecules for use in other areas of the body (for example,
enzymes made in the pancreas act in the small intestine).
Endocytosis of large particles is called phagocytosis; in our
bodies, various types of white blood cells ingest foreign particles
and bacteria by phagocytosis. Endocytosis of small particles is
called pinocytosis; an example of pinocytosis is the absorption of
small nutrient particles into the small~intestine.

Exocytosis and endocytosis operations were considered in terms of
process algebra by Cardelli \cite{Cardelli05}, with careful
biological motivation and formulation, while in terms of membrane
computing, by Cardelli and P\u aun \cite{Cardelli06}.

\begin{figure}[ht]
\begin{center}
\includegraphics[scale=0.55]{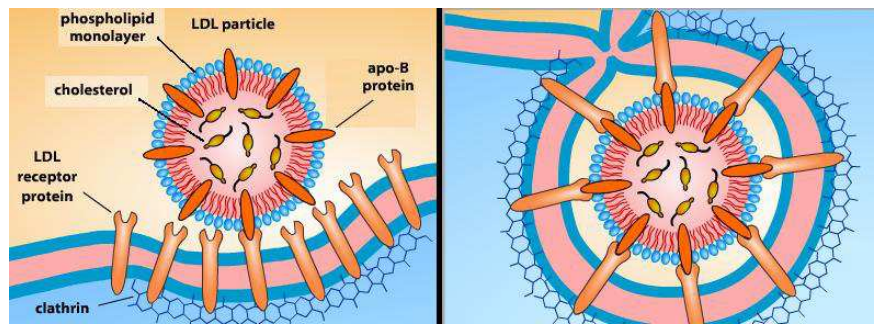}
\end{center}
\vspace{-4ex}\caption{Receptor-Mediated Endocytosis
\cite{lifepage}}\label{figure:receptor}
\end{figure}

We analyze the processes of endocytosis and exocytosis in order to
define appropriate operations for mobile membranes. In {\it
receptor-mediated endocytosis}, specific reactions at the cell
surface trigger the uptake of specific molecules \cite{lifepage}. We
present this process by an example. In such an endocytosis, a cell
takes in a particle of low-density lipoprotein (LDL) from the
outside. To do this, the cell uses receptors which specifically
recognize and bind to the LDL particle. An LDL particle contains one
thousand or more cholesterol molecules at its core. A monolayer of
phospholipids surrounds the cholesterol and it is embedded with
proteins called apo-B. These apo-B proteins are specifically
recognized by receptors in the cell membrane. The receptors in the
coated pit bind to the apo-B proteins on the LDL particle. The pit
is re-enforced by a lattice like network of proteins called
clathrin. Additional clathrin molecules are then added to the
lattice; eventually engulfing the LDL particle entirely.

\begin{figure}[ht]
\begin{center}
\includegraphics[scale=0.65]{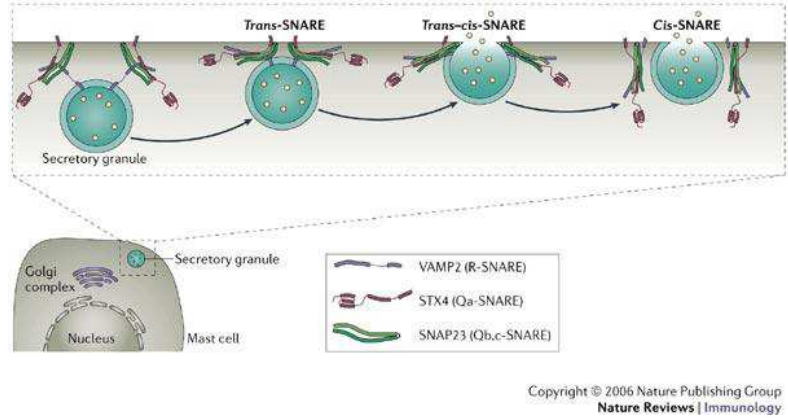}
\end{center}
\vspace{-4ex}  \caption{SNARE-Mediated
Exocytosis}\label{figure:exocytosis}
\end{figure}

{\it SNARE-mediated exocytosis} is the movement of materials out of
a cell via vesicles \cite{Albers07}. SNARES (Soluble NSF Attachment
Protein Receptor)) located on the vesicles (v-SNARES) and on the
target membranes (t-SNARES) interact to form a stable complex which
holds the vesicle very close to the target~membrane.

Endocytosis and exocytosis are modelled by mobile membranes
\cite{KrishnaPaun05}. Based on the previous examples (Figure
\ref{figure:receptor} and Figure \ref{figure:exocytosis}) where the
endocytosis is triggered by the ``agreement'' between specific
receptors and LDL particles and exocytosis by the agreement of
SNARES, we introduced in \cite{Aman09} the {\it mutual} mobile
membranes. In systems of mutual mobile membranes, any movement takes
place only if the involved membranes agree on the movement, and this
agreement is described by means of objects $a$ and co-objects
$\overline{a}$ present in the membranes involved in such a movement.
An object $a$ marks the active part of the movement, and an object
$\overline{a}$ marks the passive part. The duality relation is
distributive over a multiset, namely $\overline{u}=\overline{a_1}
\ldots \overline{a_n}$ for $u=a_1 \dots a_n$. The motivation for
introducing the systems of mutual mobile membranes comes both from
biology (e.g., receptor-mediated endocytosis), and from theoretical
computer science, namely for defining models closer to the
biological reality.

For an alphabet $V= \{a_1, \dots, a_n\}$, we denote by $V^*$ the set
of all strings over $V$; $\lambda$ denotes the empty string and
$V^+=V^*\backslash\{\lambda\}$. A multiset over $V$ is represented
by a string over $V$ (together with all its permutations), and each
string precisely identifies a multiset.

\begin{definition}
\label{definition:mutual_mobile_membranes} A system of $n \geq 1$
{\rm mutual mobile membranes} is a construct

\centerline{$\prod=(V,H,\mu,w_1,\ldots,w_n,R,i_O)$}

\noindent where:
\begin{enumerate}
\item $V$ is an alphabet (its elements are called {\rm objects});

\item $H$ is a finite set of {\rm labels} for membranes;

\item $\mu \subset H \times H$ describes the {\rm membrane
structure}, such that $(i,j)\in \mu$ denotes that a membrane
labelled by $j$ is contained into a membrane labelled by $i$; we
distinguish the external membrane (usually called the ``skin''
membrane) and several internal membranes; a membrane without any
other membrane inside it is said to be elementary;

\item $w_1,\ldots,w_n\in V^*$ are {\rm
multisets of objects} placed in the $n$ regions of $\mu$;

\item $i_O$ is the output membrane;

\item $R$ is a finite set of {\rm developmental rules} of the
following forms:
\begin{flushright} {\sf mutual endocytosis}\end{flushright}\vspace{-2ex}

\begin{enumerate}

\item $[uv]_h [\overline{u}v']_m \rightarrow [\;[w]_h w']_m$
for $h,m \in H, u, \overline{u}\in V^+,  v,v',w,w' \!\! \in V^*$;

    An elementary membrane labelled $h$ enters the adjacent membrane
    labelled $m$ under the control of the multisets of objects $u$ and
    $\overline{u}$. The labels $h$ and $m$ remain unchanged during
    this process; however the multisets of objects $uv$ and
    $\overline{u}v'$ are replaced with the multisets of objects $w$ and
    $w'$, respectively.

\begin{flushright} {\sf mutual exocytosis}\end{flushright}

\item $[\overline{u}v'[uv]_h]_m \rightarrow [w]_h [w']_m$
for $h,m \in H, u, \overline{u}\in V^+,  v,v',w,w'\!\!  \in V^*$;

    An elementary membrane labelled $h$ exits a membrane labelled $m$,
    under the control of the multisets of objects $u$ and
    $\overline{u}$. The labels of the two membranes remain unchanged,
    but the multisets of objects $uv$ and $\overline{u}v'$ are replaced
    with the multisets of objects $w$ and $w'$, respectively.

\end{enumerate}
\end{enumerate}
\end{definition}

\noindent The rules are applied according to the following
principles:

\begin{enumerate}
\item All rules are applied in parallel; the rules, the membranes,
and the objects are chosen nondeterministically, but in such a way
that the parallelism is maximal; this means that in each step we
apply a set of rules such that no further rule can be added to the
set.

\item The membrane $m$ from the rules of type $(a)$ and $(b)$
is said to be passive (identified by the use of $\overline{u}$),
while the membrane $h$ is said to be active (identified by the use
of $u$). In any step of a computation, any object and any active
membrane can be involved in one rule at most, while passive
membranes are not considered to be involved in the use of the rules
(hence they can be used by several rules at the same time as passive
membranes).

\item When a membrane is moved across another membrane, by
endocytosis or exocytosis, its whole contents (its objects) are
moved.

\item If a membrane exits the system (by exocytosis), then its
evolution stops.

\item All objects and membranes which do not evolve at a given step
(for a given choice of rules which is maximal) are passed unchanged
to the next configuration of the system.
\end{enumerate}

\noindent By using the rules in this way, we can describe
transitions among the configurations of the system. Some examples on
how rules are applied can be found in \cite{Aman09}.

\section{Mutual Mobile Membranes with Timers}
\label{section:timed_mutual_mobile_membranes}

The evolution of complicated real systems frequently involves
various interdependence among components. Some mathematical models
of such systems combine both discrete and continuous evolutions on
multiple time scales with many orders of magnitude. For example, in
nature the molecular operations of a living cell can be thought of
such a dynamical system. The molecular operations happen on time
scales ranging from $10^{-15}$ to $10^4$ seconds, and proceed in
ways which are dependent on populations of molecules ranging in size
from as few as approximately $10^1$ to approximately as many as
$10^{20}$. Molecular biologists have used formalisms developed in
computer science (e.g. hybrid Petri nets) to get simplified models
of portions of these transcription and gene regulation processes.
According to \cite{Lodish}:

\begin{enumerate}
\item[(i)] ``the life span of intracellular proteins varies from as
short as a few minutes for mitotic cyclins, which help regulate
passage through mitosis, to as long as the age of an organism for
proteins in the lens of the eye.''
\item[(ii)] ``Most cells in multicellular organisms $\dots$ carry
out a specific set of functions over periods of days to months or
even the lifetime of the organism (nerve cells, for example).''

\end{enumerate}

It is obvious that timers play an important role in the biological
evolution. We use an example from the immune system.

\begin{example}[\cite{Janeway01}]
T-cell precursors arriving in the thymus from the bone marrow spend
up to a week differentiating there before they enter a phase of
intense proliferation. In a young adult mouse the thymus contains
around $10^8$ to $2 \times 10^8$ thymocytes. About $5 \times 10^7$
new cells are generated each day; however, only about $10^6$ to $2
\times 10^6$ (roughly $2-4\%$) of these will leave the thymus each
day as mature T cells. Despite the disparity between the numbers of
T cells generated daily in the thymus and the number leaving, the
thymus does not continue to grow in size or cell number. This is
because approximately $98\%$ of the thymocytes which develop in the
thymus also die within the thymus.
\end{example}

Inspired from these biological facts, we add timers to objects and
membranes. We use a global clock to simulate the passage of time in
a membrane system.

\begin{definition}
\label{definition:timed_mutual_mobile_membranes}  A system of $n
\geq 1$ {\rm mutual mobile membranes with timers} is a construct

\centerline{$\Pi=(V,H,\mu,w_1, \ldots,w_n,R,T,i_O)$}

\noindent where: \begin{enumerate}
\item $V$, $H$, $\mu$, $w_1,\ldots,w_n$, $i_O$ are as in
Definition \ref{definition:mutual_mobile_membranes}.

\item $T\subseteq \{\Delta t \mid t \in \mathbb{N}\}$ is a set
of timers assigned to membranes and objects of the initial
configuration; a timer $\Delta t$ indicates that the resource is
available only for a determined period of time $t$;

\item $R$ is a finite set of {\rm developmental rules} of the
following forms:   \begin{flushright}{\sf object
time-passing}\end{flushright}\vspace{-2ex}

    \begin{enumerate}
    \item $a^{\Delta t} \rightarrow a^{\Delta (t-1)}$, for all $a \in V$
    and $t>0$

    If an object $a$ has a timer $t>0$, then its timer is
    decreased.

    \begin{flushright}{\sf object dissolution}\end{flushright}

    \item $a^{\Delta 0} \rightarrow \lambda$, for all $a \in V$

    If an object $a$ has its timer equal to $0$, then the object
    is replaced with the empty multiset $\lambda$, and so simulating
    the degradation of proteins.

    \begin{flushright}{\sf mutual endocytosis}\end{flushright}

    \item $[u^{\Delta\widetilde{t_u}}v^{\Delta\widetilde{t_v}}]^{\Delta t_h}_h
    [\overline{u}^{\;\Delta\widetilde{t_{\overline{u}}}}
    v^{'\Delta\widetilde{t_{v'}}}]^{\Delta t_m}_m
    \rightarrow [\;[w^{\Delta\widetilde{t_w}}]^{\Delta (t_h-1)}_h w^{'\Delta\widetilde
    {t_{w'}}}]^{\Delta t_m}_m$
    for $h,m \in H, u, \overline{u}\in V^+,  v,v',w,w' \!\! \in V^*$ and all
    timers are greater than $0$;

    For a multiset of objects $u$, $\widetilde{t_u}$ is a multiset
    of positive integers representing the timers of objects from $u$. An
    elementary membrane labelled $h$ enters the adjacent membrane
    labelled $m$ under the control of the multisets of objects $u$ and
    $\overline{u}$. The labels $h$ and $m$ remain unchanged during
    this process; however the multisets of objects $uv$ and
    $\overline{u}v'$ are replaced with the multisets of objects $w$ and
    $w'$, respectively. If an object $a$ from the multiset $uv$ has
    the timer $t_a$, and is preserved in the multiset $w$, then its
    timer is now $t_a-1$. If there is an object which appears in $w$
    but it does not appear in $uv$, then its timer is given
    according to the right hand side of the rule. Similar reasonings for the
    multisets $\overline{u}v'$ and $w'$. The timer $t_h$ of the active
    membrane $h$ is decreased, while the timer $t_m$ of the passive
    membrane $m$ remains the same in order to allow being involved
    in other rules.

    \begin{flushright} {\sf mutual exocytosis}\end{flushright}

    \item $[\overline{u}^{\;\Delta\widetilde{t_{\overline{u}}}}
    v^{'\Delta\widetilde{t_{v'}}}
    [u^{\Delta\widetilde{t_u}}v^{\Delta\widetilde{t_v}}]^{\Delta t_h}_h]^{\Delta t_m}_m
    \rightarrow [w^{\Delta\widetilde{t_w}}]^{\Delta (t_h-1)}_h[w^{'\Delta\widetilde
    {t_{w'}}}]^{\Delta t_m}_m$
    for $h,m \in H, u, \overline{u}\in V^+,  v,v',w,w'\!\!  \in V^*$ and all
    timers are greater than $0$;

    An elementary membrane labelled $h$ exits a membrane labelled $m$,
    under the control of the multisets of objects $u$ and
    $\overline{u}$. The labels of the two membranes remain unchanged,
    but the multisets of objects $uv$ and $\overline{u}v'$ are replaced
    with the multisets of objects $w$ and $w'$, respectively. The
    notations and the method of decreasing the timers are similar as
    for the previous~rule.

    \begin{flushright} {\sf membrane time-passing}\end{flushright}

    \item $[~]_h^{\Delta t} \rightarrow [~]_h^{\Delta (t-1)}$, for all $h \in H$

    For each membrane which did not participate as an active
    membrane in a rule of type $(c)$ or $(d)$, if its timer is $t>0$,
    this timer is decreased.

    \begin{flushright} {\sf membrane dissolution}\end{flushright}

    \item $[~]^{\Delta 0}_h \rightarrow [\delta]^{\Delta 0}_h$, for all
    $h \in H$;

    A membrane labelled $h$ is dissolved when its timer reaches 0.

    \end{enumerate}
\end{enumerate}
\end{definition}

These rules are applied according to the following principles:

\begin{enumerate}
\item All the rules are applied in parallel: in a step, the rules are
applied to all objects and to all membranes; an object can only be used by one rule and is nondeterministically chosen (there is no priority
among rules), but any object which can evolve by a rule of any form,
should evolve.

\item The membrane $m$ from the rules of type $(c)-(d)$ is said to be
passive (marked by the use of $\overline{u}$), while the membrane
$h$ is said to be active (marked by the use of $u$). In any step of
a computation, any object and any active membrane can be involved in
at most one rule, while passive membranes are not considered
involved in the use of rules (c) and (d), hence they can be used by
several rules (c) and (d) at the same time. Finally rule (e) is
applied to passive membranes and other unused membranes; this
indicates the end of a time-step.

\item When a membrane is moved across another membrane, by
endocytosis or exocytosis, its whole contents (its objects) are
moved.

\item If a membrane exits the system (by exocytosis), then its
evolution stops.

\item An evolution rule can produce the special object $\delta$ to
specify that, after the application of the rule, the membrane where
the rule has been applied has to be dissolved. After dissolving a
membrane, all objects and membranes previously contained in it
become contained in the immediately upper membrane.

\item The skin membrane has the timer equal to $\infty$, so it can
never be dissolved.

\item If a membrane or object has the timer equal to $\infty$, when
applying the rules simulating the passage of time we use the
equality $\infty-1=\infty$.
\end{enumerate}

\section{Mutual Mobile Membranes with and without Timers}
\label{section:tMM_MM}

The following results describing some relationships between systems
of mutual mobile membranes with timers and systems of mutual mobile
membranes without timers.

\begin{proposition}\label{PtotP}
For every systems of $n$ mutual mobile membranes without timers
there exists a systems of $n$ mutual mobile membrane with timers
having the same evolution and output.
\end{proposition}

\begin{proof}[Proof (Sketch)]
It is easy to prove that the systems of mutual mobile membranes with
timers includes the systems of mutual mobile membranes without
timers, since we can assign $\infty$ to all timers appearing in the
membrane structure and evolution rules.
\end{proof}

A somehow surprising result is that mutual mobile membranes with timers can be
simulated by mutual mobile membrane without timers.

\begin{proposition}\label{tPtoP}
For every systems of $n$ mutual mobile membranes with timers there
exists a systems of $n$ mutual mobile membrane without timers having
the same evolution and output.
\end{proposition}

\begin{proof}

We use the notation $rhs(r)$ to denote the multisets which appear in
the right hand side of a rule $r$. This notation is extended
naturally to multisets of rules: given a multiset of rules $R$, the
right hand side of the multiset $rhs(R)$ is obtained by adding the
right hand sides of the rules in the multiset, considered with their
multiplicities.

Each object $a\in V$ from a system of mutual mobile membranes with
timers has a maximum lifetime (we denote it by $lifetime(a)$) which
can be calculated as follows:

\centerline{$lifetime(a)=max(\{t\mid a^{\Delta t} \in
w_i^{\widetilde{t_i}}, 1 \leq i \leq n\}\cup\{t\mid a^{\Delta t} \in
rhs(R)\})$}

In what follows we present the steps which are required to build a
systems of mutual mobile membranes without timers starting from a
system of mutual mobile membranes with timers, such that both
provide the same result and have the same number of membranes.

\begin{enumerate}
\item A membrane structure from a system of mutual mobile membrane with
timers
\begin{center}
\begin{tikzpicture}
\draw[thick,rounded corners=4pt] (0.0,0.5) rectangle (2.0,1.5);

\node at (0.0,0.25) {$mem^{\Delta t_{mem}}$};

\node at (1.0,1.0) {$w^{\Delta \widetilde{t}}$};
\end{tikzpicture}
\end{center}

is translated into a membrane structure of a system of mutual mobile
membranes without timers in the following way

\smallskip

\begin{center}
\begin{tikzpicture}
\draw[thick,rounded corners=4pt] (0.0,0.5) rectangle (2.5,2.5);

\node at (0.0,0.25) {$mem$};

\node at (0.6,1.7) {$w$};

\node at (1.6,1.7) {$\widetilde{b_{w\;0}}$};

\node at (1.6,1.3) {$b_{mem\;0}$};

\end{tikzpicture}
\end{center}
The timers of elements from a system of mutual mobile membranes with
timers are simulated using some additional objects in the
corresponding system of mutual mobile membranes without timers, as
we show at the next steps of the translation. The object
$b_{mem\;0}$ placed inside the membrane labelled $mem$ is used to
simulate the passage of time for the membrane. The initial multiset
of objects $w^{\Delta \widetilde{t}}$ from membrane $mem$ in the
system of mutual mobile membranes with timers is translated into the
multiset $w$ inside membrane $mem$ in the corresponding system of
mutual mobile membranes without timers together with a multiset of
objects $\widetilde{b_{w\;0}}$. The multiset $\widetilde{b_{w\;0}}$
is constructed as follows: for each object $a\in w$, an object
$b_{a\;0}$ is added in membrane $mem$ in order to simulate the
passage of time.

\item The rules $a^{\Delta t} \rightarrow a^{\Delta (t-1)}$,
$a \in V$, $0< t\leq lifetime(a)$ from the system of mutual mobile
membranes with timers can be simulated in the system of mutual
mobile membranes without timers using the following rules:

\begin{enumerate}
\item[~] $a\;b_{a\;t} \rightarrow a\;b_{a\;(t+1)}$, for all $a\in V$ and
$0 \leq t \leq lifetime(a)-1$

The object $b_{a\;t}$ is used to keep track of the units of time $t$
which have passed since the object~$a$ was created. This rule
simulates the passage of a unit of time from the lifetime of object
$a$ in the system of mutual mobile membranes with timers, by
increasing the second subscript of the object $b_{a\;t}$ in the
system of mutual mobile membranes without timers.
\end{enumerate}

\item The rules $a^{\Delta 0} \rightarrow \lambda$, $a \in V$
from the system of mutual mobile membranes with timers can be
simulated in the system of mutual mobile membranes without timers
using the following rules:

\begin{enumerate}
\item[~] $a b_{a\;t_a} \rightarrow \lambda$ for all $a\in V$ such that
$t_a=lifetime(a)$

If an object $b_{a\;t_a}$ has the second subscript equal with
$lifetime(a)$ in the system of mutual mobile membranes without
timers, it means that the timer of object $a$ is $0$ in the
corresponding system of mutual mobile membranes with timers. In this
case, the objects $b_{a\;t_a}$ and $a$ are replaced by $\lambda$,
thus simulating that the object $a$ disappears together with its
timer in the system of mutual mobile membranes with timers.
\end{enumerate}

\item The rules $[u^{\Delta\widetilde{t_u}}v^{\Delta\widetilde{t_v}}]^{\Delta t_h}_h
    [\overline{u}^{\;\Delta\widetilde{t_{\overline{u}}}}
    v^{'\Delta\widetilde{t_{v'}}}]^{\Delta t_m}_m
    \rightarrow [\;[w^{\Delta\widetilde{t_w}}]^{\Delta (t_h-1)}_h w^{'\Delta\widetilde
    {t_{w'}}}]^{\Delta t_m}_m$,
    $h,m \in H, u, \overline{u}\in V^+,  v,v',w,w' \!\! \in V^*$
    with all the timers greater than $0$, from the system of mutual
    mobile membranes with timers can be simulated in the system of mutual mobile
membranes without timers using the following rules:

\begin{enumerate}
\item[~] $[u\;\widetilde{b_{u\;t1}}v\;\widetilde{b_{v\;t2}}\;b_{h\;t3}]_h
    [\overline{u}\;\widetilde{b_{\overline{u}\;t4}}\;
    v'\widetilde{b_{v'\;t5}}\;b_{h\;t6}]_m
    \rightarrow [\;[w\;\widetilde{b_{w\;t7}}b_{h\;(t3+1)}]_h w\;
    \widetilde{b_{\overline{w'}\;t8}}\;b_{h\;(t6+1)}]_m$, where\\
    $h,m \in H, u, \overline{u}\in V^+,  v,v',w,w' \!\! \in V^*$ and
    for each $b_{aj}$ we have that $0\leq j\leq lifetime(a)-1$.

    A multiset $\widetilde{b_{u\;t1}}$ consists of all objects
    $b_{a\;j}$, where $a$ is an object from the multiset $u$.
    If an object $a$ from the multiset $uv$ has
    its timer $t_a$ and it appears in the multiset $w$, then its
    timer becomes $t_a-1$. If there is an object which appears in $w$
    but it is not in $uv$, then its timer is given
    according to the right hand side of the rule. Similar reasonings are
also true for the
    multisets $\overline{u}v'$ and $w'$. The timer of the active
    membrane $h$ is increased (object $b_{h\;t3}$ is replaced by $b_{h\;(t3+1)}$),
    while the timer of the passive membrane $m$ remains the same in order
    to allow being used in other rules.
\end{enumerate}

\item The rules $[\overline{u}^{\;\Delta\widetilde{t_{\overline{u}}}}
    v^{'\Delta\widetilde{t_{v'}}}
    [u^{\Delta\widetilde{t_u}}v^{\Delta\widetilde{t_v}}]^{\Delta t_h}_h]^{\Delta t_m}_m
    \rightarrow [w^{\Delta\widetilde{t_w}}]^{\Delta (t_h-1)}_h[w^{'\Delta\widetilde
    {t_{w'}}}]^{\Delta t_m}_m$,
    $h,m \in H, u, \overline{u}\in V^+,  v,v',w,w' \!\! \in V^*$
    with all the timers greater than $0$, from the system of mutual
    mobile membranes
with timers can be simulated in the system of mutual mobile
membranes without timers using the following rules:

\begin{enumerate}
\item[~] $[\overline{u}\;\widetilde{b_{\overline{u}\;t4}}\;
    v'\widetilde{b_{v'\;t5}}\;b_{h\;t6}
    [u\;\widetilde{b_{u\;t1}}v\;\widetilde{b_{v\;t2}}\;b_{h\;t3}]_h]_m
    \rightarrow [w\;\widetilde{b_{w\;t7}}b_{h\;(t3+1)}]_h [w\;
    \widetilde{b_{\overline{w'}\;t8}}\;b_{h\;(t6+1)}]_m$, where
    $h,m \in H, u, \overline{u}\in V^+,  v,v',w,w' \!\! \in V^*$ and
    for each $b_{aj}$ we have that $0\leq j\leq lifetime(a)-1$.

    The way these rules work is similar to the previous case.
\end{enumerate}

\item The rules $[~]_h^{\Delta t} \rightarrow [~]_h^{\Delta (t-1)}$ from the
system of mutual mobile membranes with timers can be simulated in the system of
mutual mobile membranes without timers using the following rules:

\begin{enumerate}
\item[~] $b_{h\;t} \rightarrow b_{h\;(t+1)}$ for all $h\in H$ and
$0 \leq t \leq t_h-1$.

For a membrane $h$ from the system of mutual mobile membranes with
timers, $t_h$ represents its lifetime. The object $b_{h\;t}$ is used
to keep track of the units of time $t$ which have passed from the
lifetime of the membrane $h$. This rule simulates the passage of a
unit of time from the lifetime of membrane $h$ in the system of
mutual mobile membranes with timers, by increasing the second
subscript of the object $b_{h\;t}$ in the system of mutual mobile
membranes without timers.
\end{enumerate}

\item The rules $[~]^{\Delta 0}_h \rightarrow [\delta]^{\Delta 0}_h$
from the system of mutual mobile membranes with timers
can be simulated in the system of mutual mobile membranes without
timers using the following rules:

\begin{enumerate}
\item[~] $[b_{h\;t}]_h \rightarrow [\delta]_h$ for all $h\in H$ such that
$t=t_h$

If an object $b_{h\;t}$ has the second subscript equal with $t_h$ in
the system of mutual mobile membranes without timers, it means that
the timer of membrane $h$ is $0$ in the corresponding system of
mutual mobile membranes with timers. In this case, the object
$b_{h\;t}$ is replaced by $\delta$, thus marking the membrane for
dissolution and simulating that the membrane is dissolved together
with its timer in the system of mutual mobile membranes with timers.
\end{enumerate}

\end{enumerate}

\end{proof}

We are now able to prove the computational power of systems of
mutual mobile membranes with timers. We denote by
$\mathds{N}tMM_m(mutual~ endo,mutual~exo)$ the family of sets of
natural numbers generated by systems of $m \geq 1$ mutual mobile membranes with
timers by using mutual endocytosis and
mutual exocytosis rules. We also denote by $\mathds{N}RE$ the family
of all sets of natural numbers generated by arbitrary grammars.

\begin{proposition}\label{calc}
$\mathds{N}tMM_3(mutual~endo,mutual~exo)=\mathds{N}RE$.
\end{proposition}
\begin{proof}[Proof (Sketch)]
Since the output of each system of mutual mobile membranes with
timers can be obtained by a system of mutual mobile membranes
without timers, we cannot get more than the computability power of
mutual mobile membranes without timers. Therefore, according to
Theorem 3 from \cite{Aman09}, we have that the family
$\mathds{N}tMM_3$ of sets of natural numbers generated by systems of
mutual mobile membranes with timers is the same as the family
$\mathds{N}RE$ of sets of natural number generated by arbitrary
grammars.
\end{proof}

\section{From Timed Mobile Ambients to Mobile Membranes with Timers}
\label{section:tMM_tMA}

A translation of safe mobile ambients into mobile membranes is presented in
\cite{Aman08-02}, providing also an operational correspondence between these two
formalisms such that every step in safe mobile ambients is translated into a
series of well-defined steps of mobile membranes. Since an extension with time
for mobile ambients already exists \cite{Aman07,Aman07-02,Aman08}, and one for
mobile membranes is presented in this paper, it is natural to study what is the
relationship between these two extensions: timed safe mobile ambients and
systems of mutual mobile membranes with timers.

\subsection{Timed Safe Mobile Ambients}
\label{subsection:tMA}

Ambient calculus is a formalism introduced in \cite{Cardelli98} for
describing distributed and mobile computation. In contrast with
other formalisms for mobile processes such as the $\pi$-calculus
\cite{Milner99} whose computational model is based on the notion of
{\it communication}, the ambient calculus is based on the notion of
{\it movement}. An ambient represents a unit of movement. Ambient
mobility is controlled by the capabilities {\it in, out}, and {\it
open}. Capabilities are similar to prefixes in CCS
and $\pi$-calculus \cite{Milner99}. Several variants of the ambient calculus have
been proposed by adding and/or removing features of the original
calculus \cite{Bugliesi01,Zimmer02,Levi00}. Time has been considered
in the framework of ambient calculus in
\cite{Aman07,Aman07-02,Aman08}.

We use $\mathcal{P}$ to denote the set of timed safe mobile ambients; $m,n$ for
{\it ambient names}; $a,p$ for {\it ambient~tags} ($a$ stands for {\it active}
ambients, while $p$ stands for {\it passive}~ambients), and $\rho$ as a generic
notation for both tags. We write $n^{\Delta t}[P]^{\rho}$ to denote an ambient
having the timer $\Delta t$ and the tag $\rho$; the tag $\rho$ indicates that an
ambient is active or passive. An ambient $n^{\Delta t}[P]^\rho$ represents a
bounded place labelled by $n$ in which a process $P$ is executed.

The syntax of the timed safe mobile ambients is defined in Table
\ref{tmasyntaxtable}. Process {\bf 0} is an inactive process (it does nothing).
A movement $C^{\Delta t}.P$ is provided by the capability $C^{\Delta t}$,
followed by the execution of $P$. $P\,|\,Q$ is a parallel composition of
processes $P$ and $Q$.

\begin{table}[h]
  \centering
  \caption{Syntax of Timed Safe Mobile Ambients}\label{tmasyntaxtable}
  \begin{tabular}{@{\hspace{0ex}}l@{\hspace{2ex}}l@{\hspace{2ex}}l@{\hspace{2ex}}
l@{\hspace{2ex}}l@{\hspace{2ex}}l@{\hspace{0ex}}} \hline
$n,m,\ldots$ & \quad & names &
$P,Q$ & $::\,=$ & processes\\
$C$ &  $::\,=$ & capabilities &
\quad & \quad \textbf{0} & \qquad inactivity\\
\quad & \quad $in~n$ & \qquad can enter an ambient $n$&
\quad & \quad $C^{\Delta t}.\,P$ &\qquad movement\\
\quad & \quad $out~n$ & \qquad can exit an ambient $n$ & \quad &
\quad $n^{\Delta t}[P]^{\rho}$ & \qquad ambient\\
\quad & \quad $\overline{in~n}$ & \qquad allows an ambient $n$ to
enter&
\quad & \quad $P\,|\,Q$ & \qquad composition\\
\quad & \quad $\overline{out~n}$ & \qquad allows an ambient $n$ to exit &
\quad &&\\
\hline
\end{tabular}
\end{table}

In timed safe mobile ambients the capabilities and ambients are used as temporal
resources; if nothing happens in a predefined interval of time, the waiting
process goes to another state. The timer $\Delta t$ of each temporal resource
indicates that the resource is available only for a determined period of
time~$t$. If $t>0$, the ambient behaves exactly as in untimed safe mobile
ambients. When the timer $\Delta t$ expires ($t=0$), the ambient $n$ is
dissolved and the process $P$ is released in the surrounding parent ambient.
When we initially describe the ambients, we consider that all ambients are
active, and associate the tag $a$ to them.

The passage of time is described by the discrete time progress functions
$\phi_\Delta$ defined over the set $\mathcal{P}$ of timed processes. This
function modifies a process accordingly with the passage of time; all the
possible actions are performed at every tick of a universal clock. The function
$\phi_\Delta$ is inspired from \cite{Berger02} and \cite{tdpi}, and it affects
the ambients and the capabilities which are not consumed. The consumed
capabilities and ambients disappear together with their timers. If a capability
or ambient has the timer equal to $\infty$ (i.e., simulating the behaviour of an
untimed capability or ambient), we use the equality $\infty-1=\infty$ when
applying the function $\phi_{\Delta}$. Another property of the time progress
function $\phi_{\Delta}$ is that the passive ambients can become active at the
next unit of time in order to participate to other reductions.

For the process $C^{\Delta t}.P$ the timers of $P$ are activated
only after the consumption of capability $C^{\Delta t}$ (in at most
$t$ units of time). Reduction rules (Table \ref{reductiontable})
show how the time progress function $\phi_{\Delta}$ is used.

\begin{definition}\label{phideltadef}
{(Global time progress function) } We define $\phi _\Delta
:\mathcal{P}\rightarrow\mathcal{P}$, by:

\begin{center}
\begin{math}
\phi _{\Delta} (P) = \left\{
{\begin{tabular}{@{\hspace{0ex}}l@{\hspace{2ex}}l@{\hspace{0ex}}}
$C^{\Delta (t-1)}.\,R$ & if $P = C^{\Delta t}.\,R$, $t>0$\\
$R$  & if $P = C^{\Delta t}.\,R$, $t = 0$ \\
$\phi _{\Delta}(R)\ |\ \phi _{\Delta}(Q)$ & if $P =
R\,|\,Q$\\
$n^{\Delta (t-1)}[\phi _{\Delta}(R)]^a$ & if
$P=n^{\Delta t}[R]^\rho$, $t>0$\\
$R$ & if $P=n^{\Delta t}[R]^\rho$, $t = 0$\\
$P$ & if $P={\bf 0}$\\
\end{tabular}} \right.
\end{math}
\end{center}
\end{definition}

Processes can be grouped into equivalence classes by an equivalence relation
$\Xi$ called structural congruence which provides a way of rearranging
expressions so that interacting parts can be brought together. We denote by
$\not\hspace{-2pt}\dashrightarrow$ the fact that none of the rules from Table
\ref{reductiontable}, except the rule {\bf (R-TimePass)} can be applied. The
evolution of timed safe mobile ambients is given by the following reduction
rules:

\begin{table}[h]
  \centering
  \caption{Reduction rules}\label{reductiontable}

\begin{tabular}{@{\hspace{1ex}}l@{\hspace{2ex}}l@{\hspace{0ex}}}
\hline

\begin{tabular}{c} \quad\\ \textbf{(R-In)}\\ \quad\\\end{tabular} &
\begin{tabular}{@{\hspace{2ex}}c@{\hspace{2ex}}}
$-$\\
\hline $n^{\Delta t1}[in^{\Delta t2}m.P\,|\,Q]^a \,|\, m^{\Delta
t3}[\overline{in~m}^{\Delta t4}.R]^\rho \dashrightarrow m^{\Delta
t3}
[n^{\Delta t1}[P\,|\,Q]^p \,|\,R]^\rho$\\
\end{tabular}\\

\begin{tabular}{c} \quad\\ \textbf{(R-Out)}\\ \quad\\\end{tabular} &
\begin{tabular}{@{\hspace{2ex}}c@{\hspace{2ex}}}
$-$\\ \hline $m^{\Delta t3}[n^{\Delta t1}[out^{\Delta
t2}m.P\,|\,Q]^a\,|\,\overline{out~m}^{\Delta t4}.R]^\rho
\dashrightarrow
n^{\Delta t1}[P\,|\,Q]^p\,|\,m^{\Delta t3}[R]^\rho$\\
\end{tabular}\\

\textbf{(R-Amb)} &
\begin{tabular}{@{\hspace{2ex}}c@{\hspace{2ex}}}
$P \dashrightarrow Q$\\
\hline
$n^{\Delta t}[P]^\rho \dashrightarrow n^{\Delta t}[Q]^\rho$\\
\end{tabular}
\ \textbf{(R-Par1)}
\begin{tabular}{@{\hspace{1ex}}c@{\hspace{1ex}}}
$P \dashrightarrow Q$\\
\hline
$P\,|\,R \dashrightarrow Q\,|\,R$\\
\end{tabular}\\

\textbf{(R-Par2)} &
\begin{tabular}{@{\hspace{2ex}}c@{\hspace{2ex}}}
$P \dashrightarrow Q$, $P'\dashrightarrow Q'$\\
\hline
$P\,|\,P' \dashrightarrow Q\,|\,Q'$\\
\end{tabular} \hspace{2ex} \textbf{(R-Struct)}
\begin{tabular}{@{\hspace{2ex}}c@{\hspace{2ex}}}
$P' \Xi P,~P \dashrightarrow Q,~Q \Xi Q'$\\
\hline
$P'\dashrightarrow Q'$\\
\end{tabular}\\

\textbf{(R-TimePass)} &
\begin{tabular}{@{\hspace{2ex}}c@{\hspace{2ex}}}
$M \not\hspace{-3pt}\dashrightarrow$\\
\hline
$M \dashrightarrow \phi_{\Delta}(M)$\\
\end{tabular}\\
\hline
\end{tabular}
\end{table}

In the rules {\bf (R-In)}, {\bf (R-Out)} ambient $m$ can be {\it
passive} or {\it active}, while the ambient $n$ is {\it active}. The
difference between {\it passive} and {\it active} ambients is that
the {\it passive} ambients can be used in several reductions in a
unit of time, while the {\it active} ambients can be used in at most
one reduction in a unit of time, by consuming their capabilities. In
the rules {\bf (R-In)} and {\bf (R-Out)} the {\it active} ambient $n$
becomes {\it passive}, forcing it to consume only one capability in
one unit of time. The ambients which are tagged as {\it passive}
become {\it active} again by applying the global time
function {\bf (R-TimePass)}.

In timed safe mobile ambients, if a process evolves by one of the
rules {\bf (R-In)}, {\bf (R-Out)}, while another one does not
perform any reduction, then rule {\bf (R-Par1)} should be applied.
If more than one process evolves in parallel by applying one of the
rules {\bf (R-In)}, {\bf (R-Out)}, then the rule {\bf (R-Par2)}
should be applied. We use the rule {\bf (R-Par2)} to compose
processes which are active, and the rule {\bf (R-Par1)} to compose
processes which are active and passive.

\subsection{Translation}
\label{subsection:translation}

We denote by $\mathcal{M}(\Pi)$ the set of configurations obtained
along all the possible evolution of a system $\Pi$ of mutual mobile
membranes with timers.

\begin{definition}
For a system $\Pi$ of mutual mobile membranes with timers, if $M$
and $N$ are two configurations from $\mathcal{M}(\Pi)$, we say that
$M$ reduces to $N$ (denoted by $M \rightarrow N$) if there exists a
rule in the set $R$ of $\Pi$, applicable to configuration $M$ such
that we can obtain configuration~$N$.
\end{definition}

In order to give a formal encoding of timed safe mobile ambients
into the systems of mutual mobile membranes with timers, we define
the following function:

\begin{definition}\label{definition:translation} A translation $\mathcal{T}:
\mathcal{P} \rightarrow \mathcal{M}(\Pi)$ is given by:
\begin{center}
\begin{math}
\mathcal{T}(A) = \left\{
{\begin{tabular}{@{\hspace{0ex}}l@{\hspace{2ex}}l@{\hspace{0ex}}}
    $C^{\Delta t} \mathcal{T}(A_1)$& if $A=C^{\Delta t}.\,A_1$ \\
    $[\;\mathcal{T}_1(A_1)\;]^{\Delta t}_n$ & if $A=n^{\Delta t}[\;A_1\;]^\rho$\\
    $\mathcal{T}_1(A_1)\mathcal{T}_1(A_2)$ & if $A = A_1\,|\,A_2$\\
    $\lambda$ & if $A={\bf 0}$
\end{tabular}} \right.
\end{math}
\end{center}

where the system $\Pi$ of mutual mobile membranes with timers is
constructed as follows:

\centerline{$\Pi=(V,H,\mu,w_1, \ldots,w_n,R,T,i_O)$}

as follows:

\begin{itemize}
\item $n \geq 1$ is the number of ambients from $A$;

\item $V$ is an alphabet containing the $C$ objects
from $\mathcal{T}(P)$;

\item $H$ is a finite set of labels containing the labels of ambients
from $A$;

\item $\mu \subset H \times H$ describes the membrane
structure, which is similar with the ambient structure of $A$;

\item $w_i\in V^*$, $1\leq i \leq n$ are multisets of objects which
contain the $C$ objects from $\mathcal{T}(A)$ placed inside membrane
$i$;

\item $T\subseteq \{\Delta t \mid t \in \mathbb{N}\}$ is a multiset
of timers assigned to each membrane and object; the timer of each
ambient or capability from $A$ is the same in the corresponding
translated membrane or object;

\item $i_O$ is the output membrane - can be any membrane;

\item $R$ is a finite set of {\rm developmental rules}, of the
following forms:

\begin{enumerate}
\item $[in^{\Delta t2}m]_n^{\Delta t1} \,|\, [\overline{in~m}^{\Delta
t4}]_m ^{\Delta t3} \rightarrow [[~]_n^{\Delta t1} ]_m^{\Delta t3}$,
for all $n,m\in H$ and all $in~m,\overline{in~m}\in V$

\item $[[out^{\Delta
t2}m]_n^{\Delta t1}\,|\,\overline{out~m}^{\Delta t4}]_m^{\Delta t3}
\rightarrow [~]_n^{\Delta t1}\,|\,[~]_m^{\Delta t3}$, for all
$n,m\in H$ and all $out~m,\overline{out~m}\in V$
\end{enumerate}
\end{itemize}
\end{definition}

When applying the translation function we do not take into account
the tag $\rho$, since in mobile membranes a membrane is active or
passive depending on the rules which are applied in an evolution
step and we do not need something similar to ambients tags.

\begin{proposition}\label{proposition:PQ}
If $P$ is a timed safe mobile ambient such that $P \rightarrow Q$,
then there exists a system $\Pi$ of mutual mobile membranes with
timers and two configurations $M,N\in\mathcal{M}(\Pi)$, such that
$M=\mathcal{T}(P)$, $M \rightarrow N$ and $N=\mathcal{T}(Q)$.
\end{proposition}
\begin{proof}[Proof (Sketch)]

The construction of $\Pi$ is done following similar steps as in
Definition \ref{definition:translation}.

If $P \dashrightarrow Q$, then there exists a rule in the set of
rules $R$ of $\Pi$ such that $M \rightarrow N$ and
$N=\mathcal{T}(Q)$.
\end{proof}

\begin{proposition}\label{proposition:MN}
If $P$ is a timed safe mobile ambient, $\Pi$ is a system of mutual
mobile membranes with timers and $M, N\in\mathcal{M}(\Pi)$ are two
configurations, with $M=\mathcal{T}(P)$ and $M \rightarrow N$, then
there exists a timed safe mobile ambient $Q$ such that
$N=\mathcal{T}(Q)$.
\end{proposition}
\begin{proof}[Proof (Sketch)]
The system $\Pi$ of mutual mobile membranes with timers is
constructed in the same way as in Definition
\ref{definition:translation}. If $M \rightarrow N$ in the $\Pi$
system of mutual mobile membranes with timers, then there exist a
timed safe mobile ambient $Q$ such that $N=\mathcal{T}(Q)$.
\end{proof}

\begin{remark}
In Proposition \ref{proposition:MN} it is possible to have $P
\not\dashrightarrow Q$. Let us suppose that $P=n^{\Delta
t4}[in^{\Delta t1}m.in^{\Delta t2}k.out^{\Delta t3}s]^\rho \mid $
$m^{\Delta t6}[\overline{in~m}^{\Delta t5}]^\rho$. By translation we
obtain $M=[in^{\Delta t1}m\;in^{\Delta t2}k\;out^{\Delta
t3}s]^{\Delta t4}_n$ $[\overline{in~m}^{\Delta t5}]^{\Delta t6}_m$.
By constructing a system $\Pi$ of mutual mobile membrane with timers
as shown in Definition \ref{definition:translation}, we have that
$M,N \in \mathcal{M}(\Pi)$ with $M \rightarrow N$ and
$N=[[in^{\Delta t2}k\;out^{\Delta t3}s]^{\Delta t4}_n]^{\Delta
t6}_m$. For this $N$ there exists a $Q=m^{\Delta t6}[n^{\Delta
t4}[out^{\Delta t3}s.in^{\Delta t2}k]^\rho]^\rho$ such that
$N=\mathcal{T}(Q)$ but $P \not\dashrightarrow Q$.
\end{remark}

\section{Related Work}
\label{section:related_work}

There are some papers using time in the context of membrane
computing. However time is defined and used in a different manner
than in this paper. In \cite{Cavaliere05} a timed P system is
introduced by associating to each rule a natural number representing
the time of its execution. Then a P system which always produces the
same result, independently from the execution times of the rules, is
called a time-independent P systems. The notion of time-independent
P systems tries to capture the class of systems which are robust
against the environment influences over the execution time of the
rules of the system. Other types of time-free systems are considered
in \cite{Cavaliere06,Cavaliere05-02}.

Time of the rules execution is stochastically determined in
\cite{Cavaliere08}. Experiments on the reliability of the
computations have been considered, and links with the idea of
time-free systems are also discussed.

Time can also be used to ``control'' the computation, for instance
by appropriate changes in the execution times of the rules during a
computation, and this possibility has been considered in
\cite{Cavaliere06-03}. Moreover, timed P automata have been proposed
and investigated in \cite{Barbuti09}, where ideas from timed
automata have been incorporated into timed P systems.

Frequency P systems has been introduced and investigated in
\cite{Molteni08}. In frequency P systems each membrane is clocked
independently from the others, and each membrane operates at a
certain frequency which could change during the execution. Dynamics
of such systems have been investigated.

If one supposes the existence of two scales of time (an external
time of the user, and an internal time of the device), then it is
possible to implement accelerated computing devices which can have
more computational power than Turing machines. This approach has
been used in \cite{Calude04} to construct accelerated P systems
where acceleration is obtained by either decreasing the size of the
reactors or by speeding-up the communication channels.

In \cite{Cavaliere06-02,Ibarra06} the time of occurrence of certain
events is used to compute numbers. If specific events (such as the
use of certain rules, the entering/exit of certain objects into/from
the system) can be freely chosen, then it is easy to obtain
computational completeness results. However, if the length (number
of steps) are considered as result of the computation, non-universal
systems can be obtained.

In \cite{Ibarra06,Nagda06,APaun07} time is considered as the result
of the computation by using special ``observable'' configurations
taken in regular sets (with the time elapsed between such
configurations considered as output). In particular, in
\cite{Ibarra06,Nagda06} P systems with symport and antiport rules
are considered for proving universality results, and in
\cite{APaun07} this idea is applied to P systems with proteins
embedded on the membranes.

The authors of the current paper have also considered time to
``control'' the computation in two other formalisms: mobile ambients
\cite{Aman07,Aman07-02,Aman08} and distributed $\pi$-calculus
\cite{tdpi,Ciobanu07}. Timers define timeouts for various resources,
making them available only for a determined period of time. The
passage of time is given by a discrete global time progress
function.

\section{Conclusion}
\label{section:conclusion}

We introduce a new class of mobile membranes, namely the mobile
membranes with timers. Timers are assigned to each membrane and to
each object. This new feature is inspired from biology where cells
and intracellular proteins have a well defined lifetime. In order to
simulate the passage of time, we use rules of the form $a^{\Delta t}
\rightarrow a^{\Delta (t-1)}$ for objects, and $[~]_i^{\Delta t}
\rightarrow [~]_i^{\Delta (t-1)}$ for membranes. If the timer of an
object reaches $0$ then the object is consumed by applying a rule of
the form $a^{\Delta 0} \rightarrow \lambda$, while if the timer of a
membrane $i$ reaches $0$ then the membrane is marked for dissolution
by applying a rule of the form $[~]_i^{\Delta 0} \rightarrow
[\delta]_i^{\Delta 0}$. After dissolving a membrane, all objects and
membranes previously contained in it become elements of the
immediately upper membrane.

We do not obtain a more powerful formalism by adding timers to
objects and to membranes into a system of mutual mobile membranes.
According to Proposition \ref{PtotP}, Proposition \ref{tPtoP} and
Proposition \ref{calc}, systems of mutual mobile membranes with
timers and systems of mutual mobile membranes without timers have
the same computational power.

In order to relate the new class to some known formalism involving
mobility and time, we give a translation of timed safe mobile
ambients into systems of mutual mobile membranes with timers. This
encoding shows that the class of mutual mobile membranes with timers
is a powerful formalism. Such a result is related to a previous one
presented in \cite{Aman08-02}, where it is proved an operational
correspondence between the safe mobile ambients and the systems of
mutual mobile membranes.

\bibliographystyle{eptcs}

\begin{thebibliography}{1}

\bibitem{Albers07}
B. Alberts, A. Johnson, J. Lewis, M. Raff, K. Roberts, P. Walter.
\newblock {\it Molecular Biology of the Cell - Fifth Edition}.
\newblock Garland Science, Taylor \& Francis Group, 2008.

\bibitem{Aman07}
B. Aman, G. Ciobanu.
\newblock Timers and Proximities for Mobile Ambients.
\newblock {\it Lecture Notes in Computer Science}, vol.4649, 33--43,
2007.

\bibitem{Aman07-02}
B. Aman, G. Ciobanu.
\newblock Mobile Ambients with Timers and Types.
\newblock {\it Lecture Notes in Computer Science}, vol.4711, 50--63,
2007.

\bibitem{Aman08}
B. Aman, G. Ciobanu.
\newblock Timed Mobile Ambients for Network Protocols.
\newblock {\it Lecture Notes in Computer Science}, vol.5048, 234--250,
2008.

\bibitem{Aman09}
B. Aman, G. Ciobanu.
\newblock Turing Completeness Using Three Mobile Membranes.
\newblock{\it Lecture Notes in Computer Science}, vol.5715, 42--55, 2009.

\bibitem{Barbuti09}
R. Barbuti, A. Maggiolo-Schettini, P. Milazzo, L. Tesei.
\newblock Timed P Automata.
\newblock {\it Electronic Notes in Theoretical Computer Science},
vol.227, 21--36, 2009.

\bibitem{Berger02}
M. Berger.
\newblock {\it Towards Abstractions for Distributed Systems}
\newblock PhD thesis, Imperial College, Department of Computing,
2002.

\bibitem{Bugliesi01}
M.~Bugliesi, G.~Castagna, S.~Crafa.
\newblock Boxed Ambients.
\newblock {\it Lecture Notes in Computer Science}, vol.2215,
38-63, 2001.

\bibitem{Calude04}
C.S. Calude, Gh. P\u aun.
\newblock Bio-Steps Beyond Turing.
\newblock {\it Biosystems}, vol.77(1-3), 175--194, 2004.

\bibitem{Cardelli98}
L.~Cardelli, A.~Gordon.
\newblock Mobile Ambients.
\newblock {\it Theoretical Computer Science}, vol.240(1), 170-213, 2000.

\bibitem{Cardelli05}
L. Cardelli.
\newblock Brane Calculi - Interactions of Biological Membranes.
\newblock {\it Lecture Notes in Computer Science}, vol.3082,
257--280, 2005.

\bibitem{Cardelli06}
L. Cardelli, Gh. P{\u a}un.
\newblock An Universality Result for a (Mem)Brane Calculus Based on
Mate/Drip Operations.
\newblock {\it International Journal of Foundations of Computer
Science}, vol.17(1), 49--68, 2006.

\bibitem{Cavaliere06}
M. Cavaliere, V. Deufemia.
\newblock Further Results on Time-Free P Systems.
\newblock {\it International Journal on Foundational Computer
Science}, vol.17(1), 69--89, 2006.

\bibitem{Cavaliere08}
M. Cavaliere, I. Mura.
\newblock Experiments on the Reliability of Stochastic Spiking
Neural P Systems.
\newblock {\it Natural Computing}, vol.7(4), 453--470, 2008.

\bibitem{Cavaliere05}
M. Cavaliere, D. Sburlan.
\newblock Time-Independent P Systems.
\newblock {\it Lecture Notes in Computer Science}, vol.3365,
239--258, 2005.

\bibitem{Cavaliere05-02}
M. Cavaliere, D. Sburlan.
\newblock Time and Synchronization in Membrane Systems.
\newblock {\it Fundamenta Informaticae}, vol.64(1-4),
65--77, 2005.

\bibitem{Cavaliere06-02}
M. Cavaliere, R. Freund, A.Leitsch, Gh. P\u aun.
\newblock Event-Related Outputs of Computations in P Systems.
\newblock {\it Journal of Automata, Languages and Combinatorics},
vol.11(3), 263--278, 2006.

\bibitem{Cavaliere06-03}
M. Cavaliere, C. Zandron.
\newblock Time-Driven Computations in P Systems.
\newblock {\it Proceedings of Fourth Brainstorming Week on Membrane
Computing}, 133--143, 2006.

\bibitem{Aman08-02}
G. Ciobanu, B. Aman.
\newblock On the Relationship Between Membranes and Ambients.
\newblock {\it Biosystems}, vol.91(3), 515--530, 2008.

\bibitem{vaps}
G.~Ciobanu, Gh.~P\u aun, M.J.~P\'erez-Jim\'enez (editors).
\newblock {\it Applications of Membrane Computing}, Springer, Natural
Computing Series, 2006.

\bibitem{tdpi} G. Ciobanu, C. Prisacariu. Timers for Distributed Systems.
{\it Electronic Notes in Theoretical Computer Science}, vol.164(3), 81--99, 2006.

\bibitem{Ciobanu07}
G. Ciobanu, C. Prisacariu.
\newblock Coordination by Timers for Channel-Based Anonymous
Communications.
\newblock {\it Electronic Notes in Theoretical Computer Science},
vol.175(2), 3--17, 2007.

\bibitem{Zimmer02}
D.~Hirschkoff, D.~Teller, P.~Zimmer.
\newblock Using Ambients to Control Resources.
\newblock {\it Lecture Notes in Computer Science}, vol.2421,
288-303, 2002.

\bibitem{Ibarra06}
O.H. Ibarra, A. P\u aun.
\newblock Computing Time in Computing with Cells.
\newblock {\it Lecture Notes In Computer Science}, vol.3892,
112--128, 2006.

\bibitem{Janeway01}
C.A. Janeway, P. Travers, M. Walport, M.J. Shlomchik.
\newblock {\it Immunobiology - The Immune System in Health and
Disease}. Fifth Edition.
\newblock Garland Publishing, 2001.

\bibitem{KrishnaPaun05}
S.N. Krishna, Gh. P{\u a}un.
\newblock P Systems with Mobile Membranes.
\newblock {\it Natural Computing}, vol.4(3), 255--274, 2005.

\bibitem{Levi00}
F.~Levi, D.~Sangiorgi.
\newblock Controlling Interference in Ambients.
\newblock {\it Principles of Programming Languages}, 352-364, 2000.

\bibitem{Lodish}
H. Lodish, A. Berk, P. Matsudaira, C. Kaiser, M. Krieger, M. Scott,
L. Zipursky, J. Darnell.
\newblock {\it Molecular Cell Biology}. Fifth Edition, 2003.


\bibitem{Milner99}
R. Milner.
\newblock {\it Communicating and Mobile Systems: the
$\pi$-calculus}.
\newblock Cambridge University Press, 1999.

\bibitem{Molteni08}
D. Molteni, C. Ferretti, G. Mauri.
\newblock Frequency Membrane Systems.
\newblock {\it Computing and Informatics}, vol.27(3), 467--479, 2008.

\bibitem{Nagda06}
H. Nagda, A. P\u aun, A. Rodr\'iguez-Pat\'on.
\newblock P Systems with Symport/Antiport and Time.
\newblock {\it Lecture Notes In Computer Science}, vol.4361,
463--476, 2006.

\bibitem{APaun07}
A. P\u aun, A. Rodr\'iguez-Pat\'on.
\newblock On Flip-Flop Membrane Systems with Proteins.
\newblock {\it Lecture Notes In Computer Science}, vol.4860,
414--427, 2007.

\bibitem{Paun02}
Gh.~P\u aun.
\newblock {\it Membrane Computing. An Introduction}.
\newblock Springer, 2002.

\bibitem{Petre99-02}
I. Petre, L. Petre.
\newblock Mobile Ambients and P Systems.
\newblock {\it Journal of Universal Computer Science}, vol.5(9),
588--598,~1999.

\bibitem{ppage}
Web page of the P systems: \url{http://ppage.psystems.eu}.

\bibitem{lifepage}
Web page http://bcs.whfreeman.com/thelifewire.

\end{thebibliography}

\end{document}